\documentclass[conference]{IEEEtran}
\usepackage{graphicx}
\usepackage{url}
\usepackage{listings,soul,comment} 
\usepackage[
singlelinecheck=false 
]{caption}
\usepackage{color} 
\usepackage{float}
\definecolor{mygreen}{rgb}{0,0.6,0}
\definecolor{mygray}{rgb}{0.5,0.5,0.5}
\definecolor{mymauve}{rgb}{0.58,0,0.82}
 
\definecolor{darkgray}{rgb}{.4,.4,.4}
\definecolor{purple}{rgb}{0.65, 0.12, 0.82}
 
\lstdefinelanguage{JavaScript}{
keywords={typeof, new, true, false, catch, function, return, null, catch, switch, var, if, in, while, do, else, case, break},
keywordstyle=\color{blue}\bfseries,
ndkeywords={class, export, boolean, throw, implements, import, this},
ndkeywordstyle=\color{darkgray}\bfseries,
identifierstyle=\color{black},
sensitive=false,
comment=[l]{//},
morecomment=[s]{/*}{*/},
commentstyle=\color{purple}\ttfamily,
stringstyle=\color{red}\ttfamily,
morestring=[b]',
morestring=[b]"
}
 
\begin{document}

\title{Nowhere to Hide: \\Detecting Obfuscated Fingerprinting Scripts}

\author{
  {\normalsize \rm Ray Ngan ~~ Surya Konkialla ~~ Zubair Shafiq} \\  {\normalsize \rm University of California-Davis} 
} 

\maketitle

\begin{abstract}
As the web moves away from stateful tracking, browser fingerprinting is becoming more prevalent. 
Unfortunately, existing approaches to detect browser fingerprinting do not take into account potential evasion tactics such as code obfuscation. 
To address this gap, we investigate the robustness of a state-of-the-art fingerprinting detection approach against various off-the-shelf obfuscation tools. 
Overall, we find that the combination of static and dynamic analysis is robust against different types of obfuscation. 
While some obfuscators are able to induce false negatives in static analysis, dynamic analysis is still able detect these cases. 
Since obfuscation does not induce significant false positives, the combination of static and dynamic analysis is still able to accurately detect obfuscated fingerprinting scripts. 
\end{abstract}

\section{Introduction}

\vspace{.05in} \noindent \textbf{Background \& Motivation.} 
As the arms race between online tracking and anti-tracking has escalated, trackers have moved from convention stateful cookies to stateless browser fingerprinting \cite{laperdrix2020browser}. 
Unlike cookies that are observable and can deleted by users at the client-side, a browser's fingerprint is not readily observable and non-trivial to modify. 
As mainstream web browsers move towards blocking cookie-based cross-site tracking \cite{wilander2019itp23,wood2019firefox69release}, there are concerns that trackers will adopt more privacy-invasive browser fingerprinting techniques \cite{schuh2019building}. 
The sophistication and adoption of browser fingerprinting has steadily grown over the past decade \cite{bahrami2021fp}.

\vspace{.05in} \noindent \textbf{Limitations of Prior Art.} 
The privacy community is actively working on approaches to detect browser fingerprinting. 
Disconnect conducts manual expert analysis of scripts to classify domains as fingerprinting \cite{disconnect_review}. 
Researchers have manually designed heuristics \cite{Englehardt16MillionSiteMeasurementCCS} or proposed machine learning approaches to automatically detect fingerprinting scripts based on their static code analysis and/or dynamic execution analysis \cite{bird2020actions,rizzo2021unveiling,iqbal2021fingerprinting}. 
While these approaches have been shown to be effective in detecting fingerprinting scripts, they lack any explicit consideration of the adversarial threat model where fingerprinters might use obfuscation to evade detection. 
This is not a hypothetical statement---a prior study of JavaScript obfuscation found that many of the obfuscated scripts are used for browser fingerprinting \cite{skolka2019anything}. 
Obfuscation not only makes a script's code challenging to manually analyze, it is also expected to impede static and dynamic analysis techniques \cite{sarker2020hiding}. 
To the best of our knowledge, prior work on fingerprinting detection lacks an evaluation of robustness to obfuscation techniques.

\vspace{.05in} \noindent \textbf{Proposed Approach.}
In this paper, we aim to evaluate the robustness of state-of-the-art fingerprinting detection approaches against code obfuscation. 
However, it is challenging to conduct this evaluation in the wild because JavaScript obfuscation on the web is uncommon or limited to simpler obfuscation techniques \cite{skolka2019anything,sarker2020hiding}. 
Thus, any in the wild evaluation of fingerprinting detection approaches would be biased towards non-obfuscated scripts. 
To address this challenge, we design and implement a testbed that is able to seamlessly replace any target script with its obfuscated counterpart.
The testbed allows us to evaluate the robustness of fingerprinting detection approaches against JavaScript obfuscation tools in a controlled manner. 
The testbed (1) collects scripts from a web page using an instrumented browser, (2) obfuscates the scripts using off-the-shelf JavaScript obfuscation tools, (3)
replaces the original script with its obfuscated version during the next page load, and (4) evaluates the effectiveness of fingerprinting detection \cite{iqbal2021fingerprinting}. 
By comparing the output of the control (original) and the treatment (obfuscated) group, we are able to systematically evaluate the robustness of fingerprinting detection against different obfuscation tools.

\vspace{.05in} \noindent \textbf{Results.}
We use our testbed to evaluate the accuracy of a state-of-the-art fingerprinting detector \cite{iqbal2021fingerprinting} for 21,617 scripts on 1,000 websites.
The evaluation shows that using a combined method of static analysis and dynamic analysis achieves an overall accuracy of 99.8\% against various obfuscators tested. 
We find that while some obfuscation techniques are able to evade static analysis by inducing false negatives, they are still detected by dynamic analysis. 
More notably, we find that obfuscation does not induce any significant false positives; thus, combining static analysis and dynamic analysis is still effective.

In summary, our work makes the following contributions:

\begin{enumerate}

\item We design and implement a testbed to evaluate the robustness of fingerprinting detection to code obfuscation.

\item We compare the robustness of static and dynamic analysis based models and provide insights into combining static and dynamic analysis. 

\item We test multiple off-the-shelf obfuscators and analyze the evasion success of different obfuscation techniques. 

\end{enumerate}

\noindent \textbf{Paper Organization:} The rest of this paper is organized as follows. 
Section \ref{sec: related work} provides a brief overview of prior work on browser fingerprinting and JavaScript obfuscation. 
Section \ref{sec: proposed approach} describes the design of our experimental testbed to evaluate the effectiveness of fingerprinting detection on original and obfuscated scripts.
We present the results and analysis in Section \ref{sec: experimental evaluation} before concluding in Section \ref{sec: conclusion} with an outlook to future work.

\section{Background \& Related Work}
\label{sec: related work}
 
\subsection{Browser Fingerprinting}
Browser fingerprinting is a stateless web tracking technique. 
To build a browser's fingerprint, a fingerprinting script combines information gathered from HTTP headers and Web APIs via JavaScript \cite{laperdrix2020browser}.
While this fingerprint is not guaranteed to be unique or persistent \cite{laperdrix2020browser}, browser fingerprinting provides certain advantages as compared to cookies. 
A cookie-based approach is stateful; the server typically sets the cookie and the browser sends the cookie to the server from the browser's cookie jar for all subsequent requests to the server. 
Note that a user can easily remove this cookie from the browser. 
On the other hand, browser fingerprinting is stateless in that it does not require storing any state at the client-side. 
The stateless nature of browser fingerprinting means that it is less transparent and more difficult to control (e.g., remove or modify) as compared to cookies.

The research community has been investigating browser fingerprinting techniques and their deployment in the wild over the last decade or so \cite{laperdrix2020browser,bahrami2021fp,iqbal2021fingerprinting,nikiforakis2013cookieless,faizkhademi2015fpguard,Alaca16ACSACDeviceFingerprinting,Valentino18MLFingerprintingThesis,rizzo2021unveiling,bird2020actions,Englehardt16MillionSiteMeasurementCCS,Englehardt15CookiesWWW,Olejnik17BatteryStatusIWPE}.
Prior work has manually designed heuristics to detect specific types of browser fingerprinting (e.g., canvas, font, WebGL, AudioContext, Battery Status) \cite{faizkhademi2015fpguard,Englehardt16MillionSiteMeasurementCCS,Olejnik17BatteryStatusIWPE}. 
More recently, researchers have employed static and/or dynamic analysis techniques with machine learning for automated detection of browser fingerprinting \cite{iqbal2021fingerprinting,bird2020actions,rizzo2021unveiling}.
For example, Bird et al. used semi-supervised learning to detect fingerprinting scripts based on the similarity of their execution patterns to known fingerprinting scripts \cite{bird2020actions}.

Most recently, Rizzo et al. \cite{rizzo2021unveiling} and Iqbal et al. \cite{iqbal2021fingerprinting} used supervised learning models to detect fingerprinting scripts based on static and dynamic analysis. 
Static analysis captures the content and structure of the scripts that may not be executed when visiting the sites. 
Dynamic analysis captures the execution traces that include information about API accesses by the script.
The final decision is made by combining the classification results of the static and dynamic machine learning models.
Therefore, a script is detected as fingerprinting if it is classified as such by either the static analysis model or the dynamic analysis model. 
Combining static and dynamic analysis, \textsc{Fp-Inspector} \cite{iqbal2021fingerprinting} achieved the state-of-the-art 99.9\% accuracy. 
While dynamic analysis is expected to be able to better handle the use of obfuscation by fingerprinters as compared to static analysis \cite{rizzo2021unveiling,iqbal2021fingerprinting}, it is noteworthy that prior work lacks any (systematic or anecdotal) evaluation of the impact of code obfuscation on the performance of static or dynamic analysis. 
Furthermore, prior work lacks any evaluation of whether combining static and dynamic analysis is effective at improving detection accuracy of obfuscated fingerprinting scripts.

\subsection{JavaScript Obfuscation}
Obfuscation aims to decrease the understandability of code while retaining its functionality.
The code undergoes a series of transformations that make it harder to comprehend and analyze \cite{hosseinzadeh2018diversification}. 
The types of transformations vary across different obfuscation tools, which may open-source or proprietary. 
Web developers typically use these obfuscation tools to disguise the intent of their code and prevent unauthorized copying and modification \cite{javascriptobfuscator}. 
Obfuscation is also known to be used by web-based malware to hide the malicious intent of JavaScript code.

An obfuscator usually applies several transformations, so it is important to consider the effect of each one. 
An obfuscation technique typically changes a program in one or more of the following areas: control flow, data structures, and layout \cite{hosseinzadeh2018diversification}. 
Techniques that target control flow involve reconstructing a program's functions and loops in a way that makes the program harder to follow. 
Some control flow transformations have the potential to impede dynamic analysis. 
For example, dummy code injection inserts irrelevant computations into the program, obscuring the resulting execution trace (while preserving the program's original functionality). 
Data structure transformations include variable encoding and encryption. 
%
%
Finally, layout transformations focus on decreasing information available to those reading the code. 
These include removing comments, renaming identifiers, and disrupting the formatting. 
The most effective obfuscators combine techniques from multiple categories.

Obfuscators often perform minification as well.
Minification compacts code to reduce load times and improve user experience. 
Minification tools typically remove whitespace and shorten variable/function names. 
While it has a different goal than obfuscation, minification also reduces the legibility of code, but to a lesser extent.

Recent web measurements of JavaScript obfuscation have reported that certain techniques are far more prevalent than others.
Skolka et al. \cite{skolka2019anything} built a series of machine learning classifiers to detect obfuscated/minified scripts and used it to study JavaScript obfuscation on top-100k websites. 
The first phase of their study focuses on the prevalence of obfuscation/minified among the scripts collected from these sites. 
The application of their classifiers showed that minification was far more common than obfuscation. 
Specifically, they found that minification was used by 38\% of the analyzed scripts while less than $<$ 1\% of the encountered scripts were obfuscated.
The authors then analyzed these obfuscated scripts in the next phase of their study. 
They created classifiers to detect obfuscation done by each of the following tools: javascript-obfuscator, DaftLogic, jfogs, javascriptobfuscator.com, and JSObfu.
DaftLogic obfuscation was detected in nearly 90\% of the obfuscated scripts. 
The authors noted that DaftLogic is the only obfuscator among these five that encodes the entire code using \texttt{eval}. 
The authors concluded that \texttt{eval} based obfuscation is the most commonly used obfuscation technique among the detected obfuscated scripts.

More recently, Sarker et. al \cite{sarker2020hiding} leveraged the discrepancy between static and dynamic analysis outputs to detect code obfuscation. 
Using this approach, they studied JavaScript obfuscation on the top-100k website. 
They also analyzed popular JavaScript obfuscation techniques in the wild. 
The authors concluded that \texttt{eval} is no longer the most popular obfuscation technique on the web. 
They further identified five other commonly used obfuscation techniques.
The most common technique in this group is ``Functionality Map", which involves taking all of the script's invocations in string literal form and placing them in an array. 
In some cases, the array may be rotated. 
An ``accessor" function may also be used rather than direct array indices; this function generates the invocation's index based on the given argument(s).
The obfuscation technique shown in Figure ~\ref{fig:jsObfusExample} is a simplified variation of a functionality map. 
``Table of Accessors" uses a decoder function that maps encoded strings to function names and property accesses. 
Each function and property access is associated with a specific call to the decoder function. 
All of these calls are placed into an array and accessed accordingly.
The technique ``Coordinate Munging" uses wrapper functions to perform API calls and property accesses; each invocation is associated with an identifier that must be passed to the wrapper function. 
Similarly, ``Classic String Constructor" uses decoder functions to map a given number to a string literal. 
``Switch-blade Function" involves a switch-case function, which returns a wrapper function that corresponds to the requested invocation. 
The switch statement may further contain nested switch statements.

\begin{figure*}[!t]
\centerline{\includegraphics[scale=.55]{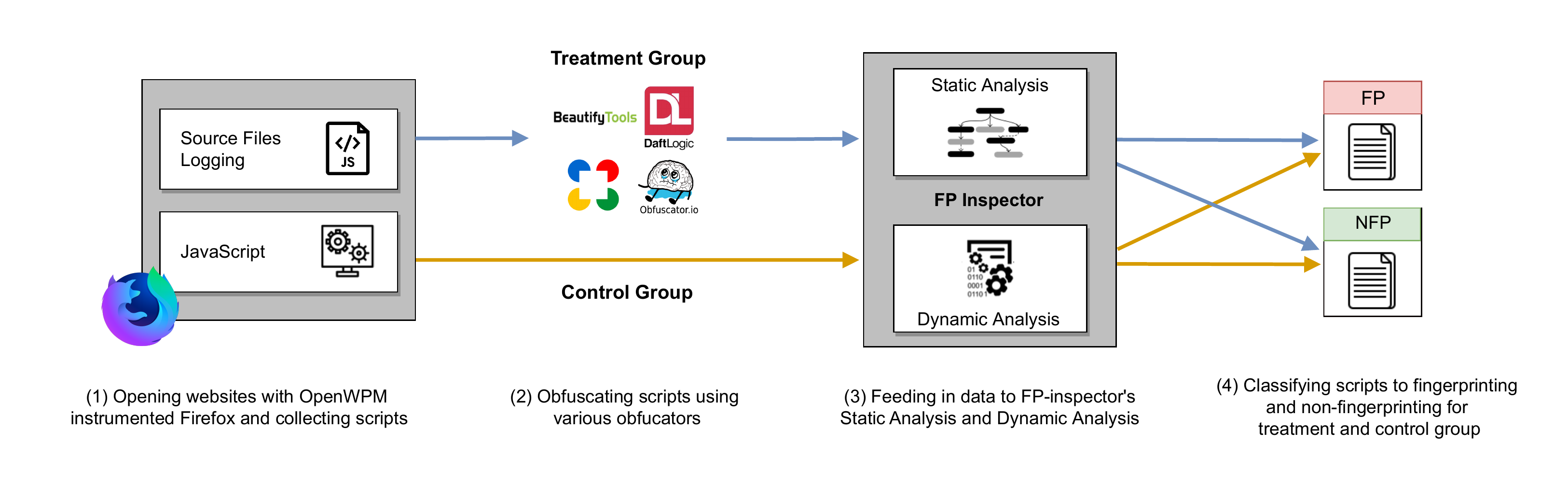}}
\caption{Overview of our evaluation approach: (1) We crawl web pages with an extended version of OpenWPM that extracts JavaScript source files and their execution traces. (2) The original JavaScript source files serve as the control group. To generate the treatment group, we pass the JavaScript source files to different obfuscators. (3) We pass the both the control group and treatment group to \textsc{Fp-Inspector}'s static analysis and dynamic analysis components. (4) We take \textsc{Fp-Inspector}'s results and identify which classifications varied across the control group and treatment group.}
\label{fig}
\end{figure*}

\section{Proposed Approach}
\label{sec: proposed approach}

\subsection{Overview}
We aim to investigate the impact of different JavaScript obfuscation techniques on the accuracy of static and dynamic analysis based fingerprinting detection approaches. 
To this end, we design and implement a testbed to collect scripts on a webpage using an instrumented browser, obfuscate them using various off-the-shelf JavaScript obfuscation tools, replace the original scripts with their obfuscated versions, and evaluate \textsc{FP-Inspector}'s static and dynamic analysis classifiers. 
First, we use OpenWPM \cite{openwpm} to crawl through a list of web pages and collect the sources of JavaScript code snippets. 
Second, we use various off-the-shelf obfuscators to obfuscate the collected scripts. 
The original, presumably non-obfuscated, scripts will be used as our control group and obfuscated scripts will serve as our treatment group.
Third, we reload the web pages in both control and treatment (i.e., replace original scripts with their obfuscated versions) settings.
Finally, we evaluate static analysis and dynamic analysis classifiers of \textsc{FP-Inspector} \cite{iqbal2021fingerprinting} in both control and treatment groups.

\subsection{Script collection}
We use OpenWPM to automatically crawl web pages and collect scripts from them. 
Our OpenWPM instrumentation is able to collect the embedded scripts that are referenced in the HTML using the ``src'' (source) attribute in \texttt{<script>} tags. 
However, our instrumentation is not able to capture inlined scripts that are directly embedded in the main HTML using the \texttt{<script>} tag.\footnote{Note that Iqbal et al. \cite{iqbal2021fingerprinting} reported that inlined fingerprinting scripts are relatively uncommon. Less than 3\% of the detected fingerprinting scripts in their dataset were inlined. }
This set of scripts serve as our control group.

\subsection{Script obfuscation}
We then obfuscate these code snippets using a set of off-the-shelf JavaScript obfuscation tools. 
Since we need to obfuscate a large number of scripts, we implemented an automation script for each obfuscator. 
Some obfuscators have a command line interface which makes the automation. 
Other obfuscators that have a web-based interface are automated using Selenium \cite{selenium}. 
This obfuscated set of scripts serves as the treatment group.

\subsection{\textsc{Fp-Inspector}}
We briefly describe the details of \textsc{Fp-Inspector}'s static and dynamic analysis models \cite{iqbal2021fingerprinting} and how we use them in our testbed. 

\subsubsection{Static Analysis}
The static analysis component of \textsc{Fp-Inspector} focuses on the content and structure of the JavaScript code snippet rather than its execution. 
Given a script, \textsc{Fp-Inspector} begins by building an Abstract Syntax Tree (AST), where the nodes represent keywords, identifiers, and literals and the edges represent the dependencies between them. 
\textsc{Fp-Inspector} then traverses the AST and extracts features. 
A decision tree model uses these features to classify the script as fingerprinting or non-fingerprinting.

To evaluate the static analysis model on control and treatment groups, we run static analysis on a set of scripts and their obfuscated variants, and then analyze whether their classifications outcomes differ. 
If \textsc{Fp-Inspector} classifies a script as fingerprinting and the obfuscated version as non-fingerprinting, we conclude that obfuscation induced a false negative. 
If \textsc{Fp-Inspector} classifies a script as non-fingerprinting and the obfuscated version as fingerprinting, we conclude that obfuscation induced a false positive. 
%
%

\subsubsection{Dynamic Analysis}
Different from static analysis, the dynamic analysis component of \textsc{Fp-Inspector} leverages JavaScript execution to capture the API calling frequency and the higher level semantics. 
We expect dynamic analysis to be more robust against obfuscated scripts, since it can extract relevant features from the execution traces while static analysis cannot. 
To evaluate the dynamic analysis model on control and treatment groups, we modify \textsc{Fp-Inspector}'s implementation as follows.

\begin{figure}[htbp]
\centerline{\includegraphics[scale=.45]{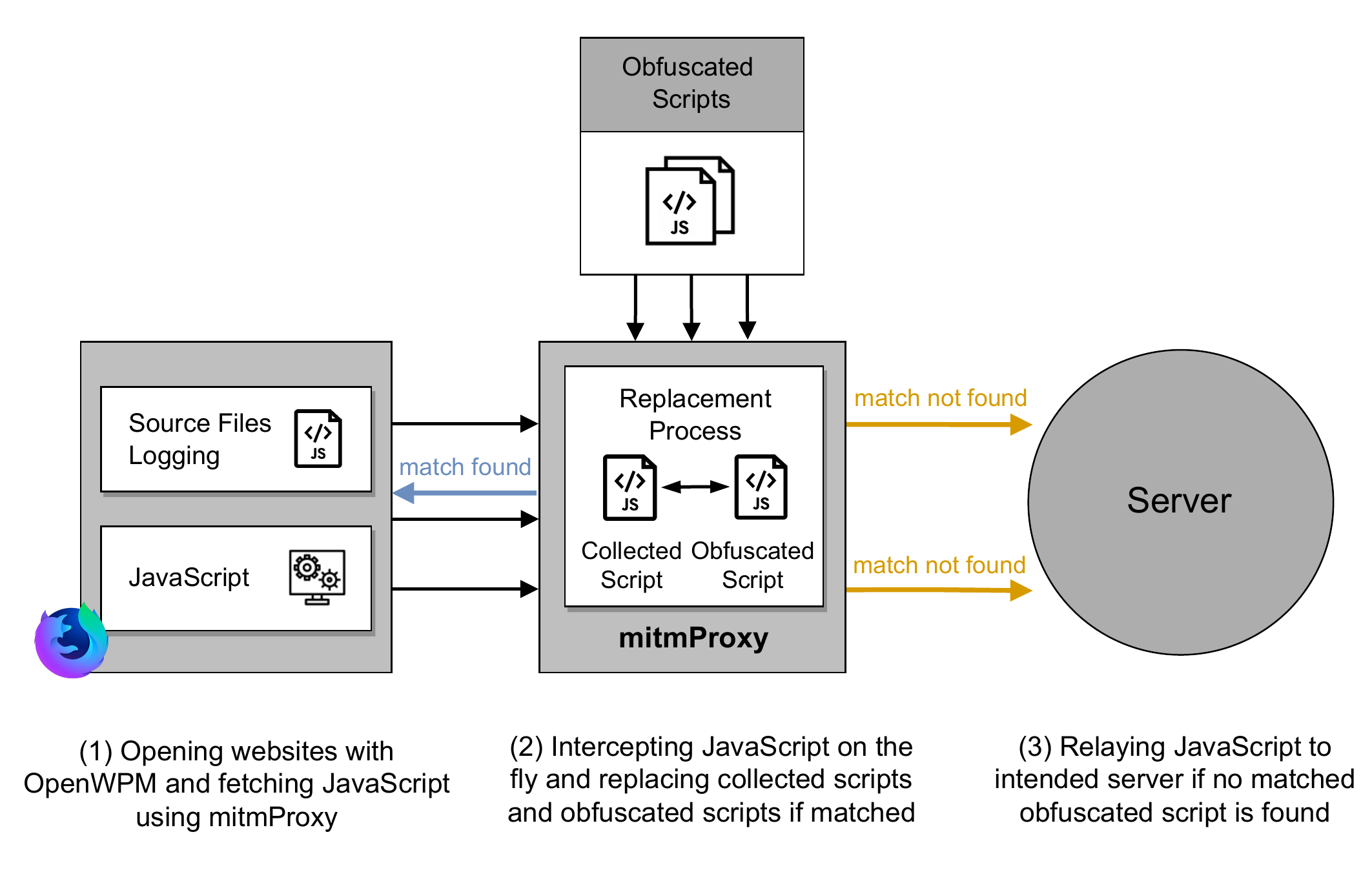}}
\caption{On-the-fly replacement of obfuscated versions of the scripts in the treatment group: 
(1) While visiting a web page using OpenWPM, we intercept HTTP requests to fetch JavaScript using the mitmProxy. 
(2) If the script URL of an intercepted HTTP request matches that of an obfuscated script, mitmProxy returns the obfuscated script as the HTTP response payload.
(3) If there is no match, then the HTTP request is relayed to the intended server to fetch the original script.}
\label{fig: on-the-fly_replacement}
\end{figure}


To seamlessly replace original scripts with their obfuscated versions during a web page load, we use a modified version of OpenWPM \cite{openwpm} and mitmProxy \cite{mitmproxy}.
Figure \ref{fig: on-the-fly_replacement} shows the process of replacing original scripts with the corresponding obfuscated scripts on the fly.
mitmProxy allows us to transparently intercept HTTP requests and modify HTTP responses in real time.
We can then map original and obfuscated scripts based on their script URL and content hash and use them to make appropriate replacements during the page reload.  
Note that we run multiple proxy instances alongside OpenWPM concurrently, one proxy instance for the control group and one proxy instance corresponding to each obfuscated variant in the treatment group, to minimize temporal differences in the control and treatment settings.
Each browser-proxy pair uses a different port; the proxy listens to requests at the specified port and replaces the scripts accordingly.

A challenge we face here is that web pages often load content dynamically; i.e., the same web page may request a different set of scripts when it is loaded in separate browser instances even if the page load time is perfectly synchronized. 
As a result, each of our proxies may intercept a different number of scripts, despite the fact that they are running concurrently.
To mitigate the issue of differing scripts across different browser-proxy instances, we decide to discard sites that could not find replacement scripts across at least 80\% of obfuscators. 
While this is not guaranteed to completely eliminate the issue, we limit our analysis to the overlapping set of scripts that is successfully obfuscated and replaced across all browser-proxy instances.

To evaluate the dynamic analysis model on control and treatment groups, we run dynamic analysis on execution traces of original scripts and their obfuscated variants, and then analyze how their classification outcomes differ.
Similar to static analysis evaluation, we record both false negatives and false positives by comparing the classification of original and obfuscated scripts. 
%


\section{Experimental Evaluation}
\label{sec: experimental evaluation}

\subsection{Data}

\subsubsection{Data Collection}
We crawl the list of web pages reported by \cite{iqbal2021fingerprinting} to contain fingerprinting scripts. 
These web pages represent an approximately 10\% subset of the top-10K sites and a random sample of 10K sites ranked between 10K and 100K \cite{fpinspector_data}. 
We re-crawl these web pages and again use \textsc{Fp-Inspector} to classify the collected scripts as fingerprinting or non-fingerprinting.
We crawled a total of 1,000 sites from \cite{fpinspector_data}. 
\textsc{Fp-Inspector} classified 543 scripts as fingerprinting and 21,074 scripts as non-fingerprinting on these sites. 
Overall,we find that 421 sites (42\%) still contain at least one fingerprinting script and 579 sites (58\%) contain only non-fingerprinting scripts.

\subsubsection{Obfuscators}
To generate the dataset of obfuscated scripts, we choose four publicly available off-the-shelf JavaScript obfuscators that implement a variety of code transformations. 
Javascript-obfuscator \cite{javascriptobfuscator} and Google Closure Compiler \cite{googleclosurecompiler} are command line tools, while DaftLogic \cite{daftlogic} and BeautifyTools \cite{beautifytools} have a web interface.

\textbf{i) javascript-obfuscator:} javascript-obfuscator is an open-source obfuscator (that also has a web-based version named obfuscator.io). 
It includes four modes that apply varying ``degrees'' of obfuscation.
Note that this obfuscator applies transformations in terms of AST nodes. 
A transformation ``threshold'' indicates the probability that that transformation will be applied to an AST node. 
For example, ``medium'' has a 75\% chance of applying control flow flattening to a particular node, while ``high'' applies this transformation to all nodes.
``Simplify'' compacts control structures by reducing the number of branches. 
``Control Flow Flattening'' replaces a program's functions and loops with an infinite loop that contains a switch statement.
``String Array'' indicates that javascript-obfuscator replaces strings with an array index (which is an implementation of the ``Functionality Map'' technique discussed in Section \ref{sec: related work}). 
Dead code injection involves adding code that does not change the execution flow.
Figure \ref{fig:js-ObfuMedium} shows an example ``Medium'' obfuscated script.


\textbf{ii) Google Closure Compiler:} Google Closure Compiler is primarily used to minify code for the sake of improving performance. 
We include this tool because it also implements obfuscation techniques. 
As shown in Figure \ref{fig:closCompSimple}, Google Closure Compiler performs whitespace and comment removal on JavaScript.
It also performs optimizations within expression and functions from JavaScript. 
This includes renaming local variables and function parameter to shorter names.


    
\textbf{iii) DaftLogic:} DaftLogic has two goals: make code more difficult to copy and paste and reduce script size to improve performance. 
This obfuscator begins by removing line breaks and comments. 
Then it applies a series of transformations, as shown in Figure \ref{fig: daftlogic}. 
We note that DaftLogic replaces keywords and identifiers with string array indices, which is an implementation of the ``Functionality Map" technique, just like ``String Array" in javascript-obfuscator. 
DaftLogic also encloses the code in an \texttt{eval}.

\textbf{iv) BeautifyTools:} 
BeautifyTools also uses \texttt{eval} and string array indices. Unlike the other obfuscators, it can also accept obfuscated input and undo the transformations. 
If the code is obfuscated with the intent of deobfuscation later, the ``Fast Decode" option can be used to quicken the process.
As shown Figures \ref{fig: beautifytools with fast decode} and \ref{fig: beautifytools without fast decode}, selecting ``Fast Decode" will define the functions beforehand.
We use the ``Fast Decode" option in our study because we see it as the option that produces more performant code. We assume that in a real use case, one would want to balance the level of obscurity and performance. The "Fast Decode" option helps find this balance.

\subsection{Evaluation Metrics}
To evaluate \textsc{Fp-Inspector}'s effectiveness, we use the classification confusion matrix. 
Since we have two classes (fingerprinting and non-fingerprinting), the confusion matrix is divided into four categories: 
True Positive (correctly detected fingerprinting script), 
True Negative (correctly detected non-fingerprinting script), 
False Positive (incorrectly detected non-fingerprinting script), and 
False Negative (incorrectly detected fingerprinting script). 
Using this confusion matrix, we calculate \textsc{Fp-Inspector}'s accuracy, false positive rate, and false negative rate for each obfuscator and separately for static and dynamic analysis classification models.

We discuss the importance of false negatives and false positives below. 
Since a fingerprinter will attempt to obfuscate its fingerprinting scripts to evade detection, its main goal is to induce false negatives.
In other words, a false negative would indicate that the obfuscation was able to successfully manipulate \textsc{Fp-Inspector} into classifying a fingerprinting script as non-fingerprinting. 
While we do not expect non-fingerprinting scripts to be obfuscated with the express purpose of evading detection, they may be obfuscated for other benign reasons. 
False positives would indicate that a non-fingerprinting obfuscated script is incorrectly classified as fingerprinting. 
These false positives would result in detection and mitigation of benign (i.e., non-fingerprinting) scripts, potentially breaking site functionality.


%

\subsection{Evaluation}


\begin{table*}
\centering
\begin{tabular}{ |c||c|c|c||c|c|c||c|c|c|c| }
    \hline
    \multicolumn{1}{|c|}{} 
    &
    \multicolumn{3}{|c|}{Combined} 
    &
    \multicolumn{3}{|c}{Static Analysis} 
    
    &                                            
    \multicolumn{3}{|c|}{ Dynamic Analysis} \\          

     \hline
     Obfuscators & FPR & FNR & Accuracy & FPR & FNR &  Accuracy & FPR & FNR & Accuracy\\
     \hline
     BeautifyTools & 0.0\% & 6.1\% & 99.5\%& 0.0\% & 100.0\%  & 93.7\%& 0.5\% & 2.3\% & 99.4\%\\
     Google Closure Compiler & 0.0\% & 0.0\% & 100.0\% & 0.0\% & 0.0\% & 100.0\% & 0.5\% & 6.8\%& 99.1\%\\
     DaftLogic & 0.0\% & 4.1\% & 99.7\%& 0.0\% & 100.0\%& 93.7\%& 0.5\% & 0.0\%& 99.5\%  \\
     Javascript-Obfuscator Default & 0.0\% & 2.0\%& 99.8\%& 0.0\% & 2.4\%& 99.8\%& 0.5\% & 6.8\% & 99.1\%  \\
     Javascript-Obfuscator Low& 0.2\% & 2.0\%& 99.7\%& 0.0\% & 4.9\%& 99.7\%& 0.5\% & 6.8\%& 99.1\% \\
     Javascript-Obfuscator Medium & 0.0\% & 2.0\%& 99.8\%& 0.2\% & 9.8\%& 99.2\%& 0.5\% & 0.0\%& 99.5\% \\
     Javascript-Obfuscator High & 0.2\% & 0.0\%& 99.8\%& 0.2\% & 2.4\% & 99.7\%& 0.7\% & 6.8\%& 98.9\%\\
     \hline
\end{tabular}

\caption{\textsc{Fp-Inspector}'s classification results in terms of False Positive Rate (FPR), False Negative Rate (FNR), and accuracy in detecting obfuscated fingerprinting scripts.}
\label{table:maintable}

\end{table*}

\textbf{Impact on obfuscators.}
Table \ref{table:maintable} presents the combined and individual results of static and dynamic models against different obfuscators. %
Note that the combined method of both static analysis and dynamic analysis produced an accuracy of over 99\% for all obfuscators. 
However, this does not necessarily mean that the obfuscation techniques did not impact \textsc{Fp-Inspector}. 
We note that BeautifyTools and DaftLogic do impact the \textsc{Fp-Inspector}'s static analysis. 
The accuracy of static analysis model for both obfuscators is reduced to around 93\%.
Crucially, the false negative rate for both BeautifyTools and DaftLogic are 100\%, meaning that all of the obfuscated fingerprinting scripts are classified as non-fingerprinting by \textsc{Fp-Inspector}'s static analysis classification model.

\begin{table}[!t]
\small
\begin{tabular}{ |c|c|c| }
    \hline
  Version & Number of Nodes & Classification\\ 
  \hline
  Original & 637 & FP\\ 
  \hline
  JS-Obfuscator (medium) & 818 & FP\\
  \hline
  DaftLogic & 93 & NONFP\\ 
  \hline
  BeautifyTools  & 93 & NONFP\\ 
  \hline
\end{tabular}
\caption{There are several instances where \textsc{Fp-Inspector} originally classified a script as fingerprinting but their obfuscated versions as non-fingerprinting. This table represents one of those instances. Note that DaftLogic's and BeautifyTools's obfuscation significantly reduces the number of AST nodes, while javascript-obfuscator's obfuscation increases the number of nodes. We generate this data using \cite{astgen}.}
\label{fig: AST nodes}
\end{table}

To understand why DaftLogic and BeautifyTools are particularly successful at evading static analysis, we investigate the construction of static analysis. 
%
%
We find that DaftLogic and BeautifyTools drastically shrink a script's AST. 
Table \ref{fig: AST nodes} shows one example of this case. 
This suggests that the efficacy of static analysis in identifying fingerprinting scripts depends on the richness of the AST.
As previously mentioned, DaftLogic and BeautifyTools are the only obfuscators in this study that use \texttt{eval} for obfuscation.
Thus, we conclude that \texttt{eval} based obfuscation, which essentially converts the code to a string form, disproportionately hinders the ability of static analysis to extract useful features. 


\textbf{Impact on static vs. dynamic analysis.}
Table \ref{table:maintable}, it also shows that dynamic analysis performs better than static analysis against all obfuscators, especially for BeautifyTools and DaftLogic. 
The overall accuracy of \textsc{Fp-Inspector}'s dynamic analysis model is around 99\%. 
This is due to the fact that dynamic analysis does not share static analysis' struggles when it comes to \texttt{eval}-based obfuscation because the former relies on code execution. 
It extracts features from the execution traces, which are not significantly impacted by obfuscation. 
While techniques like dummy code injection can sometimes alter the execution trace, \textsc{Fp-Inspector} is still be able detect the API calls made by the original code.

\textbf{Impact on combined static and dynamic analysis.}
\textsc{Fp-Inspector} \cite{iqbal2021fingerprinting} uses the OR combination of static and dynamic analysis. 
This choice, over the AND combination, is justified by the authors to overcome the false negatives of dynamic analysis due to coverage issues. 
%
This means that if either static or dynamic analysis suggests that the script is fingerprinting, then \textsc{Fp-Inspector} classifies the script as fingerprinting. 
In our experiment, we see that the OR operation of both analysis does indeed improve the performance of identifying fingerprinting scripts even the dataset is obfuscated. 
The OR approach can mitigate the false negatives of static analysis since dynamic analysis typically captures the fingerprinting scripts that are missed by static analysis.

Our results also have implications for relative importance of false positives versus false negatives. 
If either static or dynamic analysis falsely classifies a non-fingerprinting script as fingerprinting (i.e., a false positive) then using the OR operation approach in this case would hurt the overall accuracy of \textsc{Fp-Inspector}. 
In our evaluation, however, we do not see a significant number of false positives for either static or dynamic analysis. 
Therefore, based on our results, we conclude that \textsc{Fp-Inspector}'s use of the OR operation is robust to code obfuscation which is more likely to introduce false negatives than false positives.

\section{Concluding Remarks}
\label{sec: conclusion}
In this short paper, we analyzed the impact of obfuscation on the detection of fingerprinting scripts. 
We collected 21,617 scripts on 1,000 sites and obfuscated them using four publicly available obfuscators. 
We then compared the classification outcomes of the original scripts versus their obfuscated variants on \textsc{Fp-Inspector}, a state-of-the-art fingerprinting script classification approach. 
Our results showed that static analysis fails to detect fingerprinting scripts obfuscated by DaftLogic and BeautifyTools because they significantly impact their AST representation. 
However, dynamic analysis was able to still accurately classify obfuscated scripts. 
Furthermore, both static and dynamic analysis had a low false positive rate, justifying \textsc{Fp-Inspector}'s use of the OR operation in combining the outcomes of static and dynamic analysis classifiers. 
The combined static and dynamic analysis approach of \textsc{Fp-Inspector} still achieves an accuracy of more than 99\% for obfuscated scripts.

While \textsc{Fp-Inspector}'s accuracy holds against obfuscated scripts when taking the OR operation of both static and dynamic analysis classifiers, using the OR operation would not work well if an obfuscation technique were to induce a high false positive rate for either static or dynamic analysis.
While we did not observe this in this work, this might be a concern in the future as new obfuscation techniques are designed and deployed on benign scripts \cite{sarker2020hiding}. 
Future work can investigate re-training \textsc{Fp-Inspector}'s classification models on a dataset of scripts that includes obfuscated scripts. 
Such \textit{adversarial training} on a variety of different obfuscators can help build robustness against code obfuscation. 
Another option is to use obfuscation detection classifiers  \cite{Skolka19minifiedobfuscatedJS} and have separate classification pipelines for obfuscated and non-obfuscated scripts.

There are two main limitations of our experimental evaluation.
First, we encountered some scripts that could not be obfuscated. 
Some obfuscators impose strict syntax requirements (e.g., on semicolon usage).
Some scripts could not be obfuscated \textit{further} given that they were already obfuscated. 
Second, we were not able to replace some obfuscated scripts in our testbed due to the dynamism. 
Future work could mitigate this issue by using an approach of reloading each page multiple times or by looking into web page replay techniques to side-step this problem altogether.

%


\bibliographystyle{unsrt}
\bibliography{main}

\clearpage

\section{Appendix}
\noindent{Example of an original script and its different obfuscated versions.}

\begin{figure}[h]
\centering
(a) original script
\begin{lstlisting}[language=JavaScript,numbers=none]
function welcome() {
    console.log("Hi, how are you?");
}
welcome();
\end{lstlisting}
(b) obfuscated script
\begin{lstlisting}[language=JavaScript,numbers=none]
var _0xd99e=["\x48\x69\x2C\x20\x68\x6F\x77\x20\x61\x72\x65\x20\x79\x6F\x75\x3F","\x6C\x6F\x67"];function welcome(){console[_0xd99e[1]](_0xd99e[0])}welcome()
\end{lstlisting}

\vspace{-.2in}
\caption{An example JavaScript code snippet and its obfuscated version using \cite{jsobfus}.}
\label{fig:jsObfusExample}
\end{figure}

\begin{figure}[h]
\begin{lstlisting}[language=JavaScript,numbers=none]
(function(_0x5182ff,_0xd2dd0b){function _0x194a1e(_0x3fb32d,_0x16e1b4,_0x4e34a1,_0x4a7652){return _0x310c(_0x3fb32d-0x3b3,_0x16e1b4);}function _0x44770d(_0x3583bf,_0x587834,_0xa42cf0,_0x2e7719){return _0x310c(_0x2e7719-0x1a,_0xa42cf0);}var _0x39f449=_0x5182ff();while(!![]){try{var _0x47a84f=-parseInt(_0x194a1e(0x45a,0x45b,0x437,0x43f))/(0x5*0x1f3+0x1ec4+-0x2882)*(-parseInt(_0x194a1e(0x458,0x45d,0x458,0x43e))/(-0x1cc*0x3+0xb*0x5e+0x15c))+-parseInt(_0x194a1e(0x49d,0x493,0x4b8,0x4a8))/(0x562*0x7+-0x2623+0x78)*(parseInt(_0x44770d(0x10f,0x11b,0xcf,0xf4))/(-0xf9c*0x2+0xd00+-0x185*-0xc))+-parseInt(_0x194a1e(0x47c,0x454,0x497,0x4a2))/(-0xdf*-0x3+-0x6*0xb3+0xcd*0x2)
\end{lstlisting}
\vspace{-.2in}
\caption{The javascript-obfuscator's ``medium" variant of the code snippet in Figure ~\ref{fig:jsObfusExample}. This figure only shows the beginning segment of the generated output.}
\label{fig:js-ObfuMedium}
\end{figure}

\begin{figure}[h]
\begin{lstlisting}[language=JavaScript,numbers=none]
eval(function(p,a,c,k,e,d){e=function(c){return c};if(!''.replace(/^/,String)){while(c--){d[c]=k[c]||c}k=[function(e){return d[e]}];e=function(){return'\\w+'};c=1};while(c--){if(k[c]){p=p.replace(new RegExp('\\b'+e(c)+'\\b','g'),k[c])}}return p}('7 0(){6.5("4, 3 2 1?")}0();',8,8,'welcome|you|are|how|Hi|log|console|function'.split('|'),0,{}))
\end{lstlisting}
\vspace{-.2in}
\caption{The DaftLogic variant of the code snippet in Figure ~\ref{fig:jsObfusExample}.}
\label{fig: daftlogic}
\end{figure}

\begin{figure}
\vspace{-1.8in}
\begin{lstlisting}[language=JavaScript,numbers=none]
function welcome(){console.log("Hi, how are you?")}welcome();
\end{lstlisting}
\vspace{-.2in}
\caption{The Closure Compiler ``simple" variant of the code snippet in Figure ~\ref{fig:jsObfusExample}.}
\label{fig:closCompSimple}
\end{figure}

\begin{figure}
\vspace{-5.25in}
\begin{lstlisting}[language=JavaScript,numbers=none]
eval(function(p,a,c,k,e,d){e=function(c){return c};if(!''.replace(/^/,String)){while(c--){d[c]=k[c]||c}k=[function(e){return d[e]}];e=function(){return'\\w+'};c=1};while(c--){if(k[c]){p=p.replace(new RegExp('\\b'+e(c)+'\\b','g'),k[c])}}return p}('2 0(){1.3("7, 4 6 5?")}0();',8,8,'welcome|console|function|log|how|you|are|Hi'.split('|'),0,{}))
\end{lstlisting}
\vspace{-.2in}
\caption{The BeautifyTools variant of the code snippet in Figure ~\ref{fig:jsObfusExample} with the ``Fast Decode" option.}
\label{fig: beautifytools with fast decode}
\end{figure}

\begin{figure}
\vspace{-10.5in}
\begin{lstlisting}[language=JavaScript,numbers=none]
eval(function(p,a,c,k,e,d){while(c--){if(k[c]){p=p.replace(new RegExp('\\b'+c+'\\b','g'),k[c])}}return p}('2 0(){1.3("7, 4 6 5?")}0();',8,8,'welcome|console|function|log|how|you|are|Hi'.split('|')))
\end{lstlisting}
\vspace{-.2in}
\caption{The BeautifyTools variant of the code snippet in Figure ~\ref{fig:jsObfusExample} without the ``Fast Decode" option.}
\label{fig: beautifytools without fast decode}
\end{figure}

\end{document}